\documentclass[pre]{revtex4}
\usepackage{dcolumn}
\usepackage{graphicx}
\usepackage{latexsym}
\usepackage{amsmath}
\usepackage{amssymb}
\usepackage{amsfonts}
\usepackage{amsthm}
\begin{document}

\title{Quantum corrections to nonlinear ion acoustic wave with Landau damping}
\author{ Abhik Mukherjee$^1$,  Anirban Bose$^2$,  M.S. Janaki$^1$}
\affiliation{$^1$Saha Institute of Nuclear Physics,\\
Calcutta, INDIA\\
$^2$Serampore College, West Bengal, India}

\begin{abstract}
 Quantum corrections to nonlinear ion acoustic wave with Landau damping have been computed using Wigner equation 
approach. The dynamical equation governing the time development of nonlinear ion acoustic wave
with semiclassical quantum corrections is shown
to have the form of higher KdV equation which has higher order nonlinear
terms coming from quantum corrections, with the usual classical
and quantum corrected Landau damping integral terms.
 The 
conservation of total number of ions is shown from the evolution equation.  The decay rate of KdV solitary wave amplitude due to presence of Landau damping terms
 has been calculated assuming the Landau damping parameter $\alpha_1 = \sqrt{{m_e}/{m_i}}$
to be of the same order of the quantum parameter $Q = {\hbar^2}/({24 m^2 c^2_{s} L^2})$. The amplitude 
is shown to decay very slowly  with time as determined by the quantum factor  $ Q$.
\end{abstract}

 \maketitle

\section{Introduction} 
The study of plasmas, is in general limited to the domain of classical
physics where temperature is high and particle density is low. In recent years, the study of plasmas  such as  dense
astrophysical plasmas {\cite{Astro}}, laser plasmas {\cite{Laser}} as well as miniature electronic devices that are under extreme physical
conditions requires {\cite{Electronics1},\cite{Electronics2}} quantum mechanical effects to be taken into account. In such systems, the  scale length becomes comparable to the particle de Broglie wavelength rendering classical transport models unsuitable and quantum
mechanical effects to be relevant. In broad aspect there are mainly two approaches
to model quantum plasmas which are quantum hydrodynamic approach{\cite{Haas1},\cite{Manfredi}} and quantum 
kinetic approach{\cite{Bonitz}} i.e, Wigner equation  approach.
The plasma fluid equations with the inclusion of quantum diffraction and statistical pressure effects give rise to new physical
phenomena in the context of linear and nonlinear waves and instabilities. Haas\cite{Haas2} et al. have examined  quantum quasilinear  plasma turbulence using quasilinear equation derived from Wigner-Poisson system.

The quantum fluid equations being macroscopic in nature are relatively simple  and are easily accessible for nonlinear calculations. However, working with such macroscopic models  leads to loss of understanding in the situations where single particle effects  like Landau damping are important and
which can be explored by moving into a kinetic picture.
The kinetic description of  plasma possessing quantum mechanical features is provided
by the Wigner equation that can be considered as the quantum analogue of
the Vlasov equation. It describes the evolution of the
quantum mechanical phase space distribution function given by the
Wigner-Moyal distribution and can be a useful tool to look into
the microscopic nature of the system.  The Wigner function is called quasi-distribution as it can have negative values although
its velocity moments give rise to various physical variables such
as density, current etc. 
Gardner\cite{Gardner} derived  
the full three-dimensional quantum hydrodynamic (QHD) model   for the first time by a
moment expansion of the Wigner-Boltzmann equation.

 So far nonlinear problems like KdV equation
and BGK modes have been tackled successfully
in classical plasma. Recently, Lange et al.\cite{Lange}  have provided a generalization of the classical BGK modes by obtaining a solution of the stationary Wigner-Poisson equation.
In this work we have attempted to look into the quantum
KdV problem in the semi-classical limit.
For a  classical plasma {\cite{OttSudan1}} Ott and Sudan have modeled nonlinear ion acoustic wave in a kinetic picture
taking  the mass of electron into account. They obtained a KdV equation together with a Landau damping term as an evolution
equation for the  ion acoustic wave. In order to explore the quantum corrections to the nonlinear evolution of an ion acoustic wave
in presence of  Landau damping terms we have to replace the Vlasov equation by the Wigner equation.
In this article we have tried to investigate, in the semiclassical limit, the quantum corrections
to nonlinear ion acoustic  wave with Landau damping. 
We have derived a higher order KdV equation which has
higher order nonlinear quantum corrections with the usual classical Landau damping term and a term containing the quantum
corrections due to Landau damping  as the dynamical evolution equation. The equation converges to the same equation as derived by  Ott and Sudan in the classical limit i.e, when $\hbar$ tends to zero. The equation shows some features like conservation of total ion number , decay of initial waveform due to 
Landau damping etc. In the next stage we have carried out the 
perturbative approach of Bogoliubov and Mitropolsky to get the decay nature of KdV solitary wave amplitude. 
For this purpose we have assumed the Landau damping parameter $\alpha_1 $ to be of the order of  the quantum factor $Q$.
The procedure  reveals that the amplitude  decays inversely 
with the square of time depending on the factor $Q$. 

The paper is organized in the following manner. In section-II the derivation 
of the evolution equation of ion acoustic wave with the Landau damping term and the quantum corrections
is given. Some relevant properties of this higher order kdV equation are discussed in section-III. Subsection III-A discusses the  conservation of total number of ions. The Bogoliubov- Mitropolsky perturbation approach with the condition $\alpha_1 \approx Q$
and the decay nature of the KdV solitary wave is given in subsection III-B. The conclusive remarks are given in 
section-IV.

%----------------------------------------------------------------------------------------------------------------------------%
%------------------------------------------------------------------------------------------------------------------------------%
%-------------------------------------------------------------------------------------------------------------------------------%

\section{Derivation of the dynamical equation}
The Wigner distribution function is a function of the phase-space variables (x, v) and time, which,
is given by N single particle wave function $\psi_{\alpha}(x,t)$ each characterized by
a probability $P_{\alpha}$ satisfying $\sum_{\alpha=1}^N P_{\alpha}=1$. 

It is given as,
\begin{equation}  
  f(x, v, t) = \sum_{\alpha=1}^N\frac{m}{2 \pi \hbar}P_{\alpha}\int_{-\infty}^{\infty}\psi_{\alpha}^* (x+\lambda/2,t)
 \psi_{\alpha} (x-\lambda/2,t)e^\frac{ i m v \lambda}{\hbar} d\lambda,
\label{Wigner}
\end{equation}
where $m$ is the mass of the particle.
The Wigner function 
 follows the following evolution equation called the Wigner equation
  \begin{eqnarray}
  \frac{\partial f}{\partial t} + v \frac{\partial f}{\partial x} + \frac{ e m }{2 i \pi \hbar^2}
\iint[\phi(x+ \lambda/2)
- \phi(x- \lambda/2)] f(x, v', t) e^\frac{ i m \lambda (v - v')}{\hbar} d\lambda dv'=0,\label{WignerE}
 \end{eqnarray}
 where $ \hbar, \phi$ are the reduced Planck's constant and self- consistent
 electrostatic potential.
  Considering semi-classical limit,
we develop the integral upto $O(\hbar ^2)$ and neglect all higher order terms containing
$\hbar$ to obtain
\begin{equation}
\frac{\partial f}{\partial t} + v \frac{\partial f}{\partial x} + \frac{e}{m}
\frac{\partial \phi}{\partial x}\frac{\partial f}{\partial v}-
 (\frac{e \hbar^2}{24 m^3}) \frac{\partial^3 \phi}{\partial x^3}\frac{\partial^3 f}{\partial v^3}=0
\label{TWignerE}\end{equation}
%where $Q= \frac{\hbar^2}{24 m^2 v_{th}^2 L^2}$.
 We can see from (\ref{TWignerE}) that the Vlasov equation is recovered in the limit $\hbar \rightarrow 0$.

In our work, we consider a situation where ions are cold ($T_{i}=0$) and 
 electrons have finite temperature and the quantum effects are relevant for electrons only. Therefore, we consider the usual 
 fluid equations for describing the dynamics of ions and the
 Wigner equation for describing the electrons.

Hence in this case the relevant normalized system of one-dimensional equations are -
\begin{equation}
 \frac{\partial n}{\partial t} + \frac{\partial (nu)}{\partial x} = 0,
\label{continuity}
\end{equation}
which is the continuity equation for ions. The momentum conservation equation for the ions is given by
\begin{equation}
\frac{\partial u}{\partial t} + u \frac{ \partial u}{\partial x} =
 - \frac{\partial \phi}{\partial x},
\label{momentum}
\end{equation}

%,\nonumber\\
\begin{equation}
(\frac{\lambda_{D}}{L})^2\frac{\partial ^2 \phi}{\partial x^2} = n_{e} - n ,
\label{Poisson}
\end{equation}
which is the Poisson's equation appropriate for the description of
dispersive ion acoustic waves. The electron number density
if obtained as the velocity space average of the single particle
distribution function $f$
\begin{equation}
n_{e} = \int_{-\infty}^{\infty} f dv,
\label{numberdensity}
\end{equation}
that is described by the Wigner equation in the semiclassical limit 
%\nonumber\\
\begin{equation}
(\frac{m_{e}}{m_{i}})^\frac{1}{2}\frac{\partial f}{\partial t} + v \frac{\partial f}{\partial x} + 
\frac{\partial \phi}{\partial x}\frac{\partial f}{\partial v}-
 Q \frac{\partial^3 \phi}{\partial x^3}\frac{\partial^3 f}{\partial v^3}=0,
\label{Wigner},
\end{equation}
where $n_e,  n, u$ are the 
electron number density, ion number density and
 ion velocity respectively, $\lambda_{D}=\sqrt{{KT_{e}}/{4\pi n_0 e^2}}$ 
is the Debye Length, $L$ is the characteristic length for variations of $n,u,\phi,n_e,f$ and
$Q$ is the quantum parameter $= {\hbar^2}/{24 m^2 c^2_{s} L^2}$.

Here the following normalization scheme has been used :
\begin{eqnarray}
  x = \frac{\tilde{x}}{L}, t = \frac{c_{0}\tilde{t}}{L}, v = \frac{\tilde{v}}{c_s}, 
\phi = \frac{e \tilde{\phi}}{K_{B}T_{e}},
 n = \frac{\tilde{n}}{n_{0}}, f = \frac{\tilde{f}}{n_{0}}, u = \frac{\tilde{u}}{c_{0}},
\end{eqnarray}
where $c_{0}$ is the ion acoustic sound speed $= \sqrt{{K T_e}/{m_i}}$, $c_s$ is the electron thermal
velocity $= \sqrt{{K T_e}/{m_e}}$ , $n_0$ is the ambient number density of electrons (ions)
and $T_e$ is the electron temperature.

 As in case of {\cite{OttSudan1}}, here also three basic parameters enter into the problem
which are parameters due to Landau damping by electrons, measure of nonlinearity and
measure of dispersive effects. In this calculation we do not neglect the electron to ion mass ratio and
since $T_i = 0$, the Landau damping is provided solely by electrons. We consider all these three effects i.e.,
Landau damping, nonlinearity and dispersion to be small but of the same order of magnitude.

1)$\sqrt{({m_{e}}/{m_{i}})} = \alpha_1 \epsilon$, effect due to 
 Landau damping by electrons.

2)${\vartriangle n}/{n_{0}} = \alpha_2 \epsilon$, measure of the strength of nonlinearity.

3) $({\lambda_{D}}/{L})^2= 2 \alpha_3 \epsilon$, measure of strength of dispersive effects.

Here $\epsilon$ is smallness parameter. As is the usual mathematical procedure we transform our co-ordinates to a moving frame with a stretched time as
\begin{eqnarray}
 \xi = x - t,\ \tau = \epsilon t,
\label{Movframe}
\end{eqnarray}
and expand the dependent variables for small nonlinearity as 
\begin{eqnarray}
 n = 1 + \alpha_{2}\epsilon n^{(1)} + \alpha_{2}^2 \epsilon^2 n^{(2)} + ...,\nonumber\\
u =  \alpha_{2}\epsilon u^{(1)} + \alpha_{2}^2 \epsilon^2 u^{(2)} + ...,\nonumber\\
\phi =  \alpha_{2}\epsilon \phi^{(1)} + \alpha_{2}^2 \epsilon^2 \phi^{(2)} + ...,\nonumber\\
n_{e} = 1 + \alpha_{2}\epsilon n_{e}^{(1)} + \alpha_{2}^2 \epsilon^2 n_{e}^{(2)} + ...,\nonumber\\
f = f^{(0)} + \alpha_{2}\epsilon f^{(1)} + \alpha_{2}^2 \epsilon^2 f^{(2)} + ...
\label{VarExpand}
\end{eqnarray}
Considering semiclassical limit, the form of $f^{(0)}$ is chosen as
\begin{equation}
 f^{(0)}(v) = \frac{1}{\sqrt{2 \pi}}\exp{(\frac{-v^2}{2})}
\label{Maxwellian}
\end{equation}

Substituting Eqns. (\ref{Movframe}), (\ref{VarExpand}), (\ref{Maxwellian}) in (\ref{continuity})-(\ref{Wigner}) 
and equating coefficients
of $\epsilon$, $\epsilon^2$ to zero we get first and second order equations which
need to be solved.

\subsection {$\epsilon$ order calculation:}

From Eqns (\ref{continuity})-(\ref{Poisson}) we get
\begin{eqnarray}
 \frac{\partial n^{(1)}}{\partial \xi}=\frac{\partial u^{(1)}}{\partial \xi}=\frac{\partial \phi^{(1)}}{\partial \xi},
 n^{(1)} = n_{e}^{(1)}
\label{1unphi}
\end{eqnarray}
From equation (\ref{Wigner}) we get
 \begin{equation}
 v \frac{\partial f^{(1)}}{\partial \xi} = v \frac{\partial \phi^{(1)}}{\partial \xi} f^{(0)}+
Q \frac{\partial^3 \phi^{(1)}}{\partial \xi^3}(3v - v^3) f^{(0)},
\label{nonuneque1}
\end{equation}
which yields 
\begin{equation}
  \frac{\partial f^{(1)}}{\partial \xi} =  \frac{\partial \phi^{(1)}}{\partial \xi} f^{(0)}+
Q \frac{\partial^3 \phi^{(1)}}{\partial \xi^3}(3 - v^2) f^{(0)} + \lambda(\xi,\tau)\delta(v),
\label{nonuneque}
\end{equation}
where $\delta(v)$ is the Dirac delta function and $\lambda (\xi,\tau)$ is an arbitrary function of $\xi,\tau$.
Here also the problem of non-uniqueness arises as in case of {\cite{OttSudan1},\cite{AnupB}} which can be removed by
taking a $\tau$ derivative term from higher $\epsilon$ order. Thus, we write 
\begin{equation}
 (\alpha_1 \epsilon^2)\frac{\partial f_{\epsilon}^{(1)}}{\partial \tau}+
v \frac{\partial f_{\epsilon}^{(1)}}{\partial \xi} = v \frac{\partial \phi^{(1)}}{\partial \xi} f^{(0)}+
Q \frac{\partial^3 \phi^{(1)}}{\partial \xi^3}(3v - v^3) f^{(0)}, 
\label{uneque1}
\end{equation}
where the first term of (\ref{uneque1}) has been taken from order $\epsilon^3$ equation. 
Once $f_{\epsilon}^{(1)}$ is known,
$f^{(1)}$ can be determined uniquely by :
\begin{equation}
 f^{(1)} =  \lim_{\epsilon \to 0}  f_{\epsilon}^{(1)}
\end{equation}
We introduce Fourier transform in $\xi$ and $\tau$ as
\begin{equation}
\widehat{ f_{\epsilon}^{(1)}}(\omega, k)  = \frac{1}{2 \pi}\int_{\xi =-\infty}^{\infty}\int_{\tau=0}^{\infty}
 f_{\epsilon}^{(1)}(\xi,\tau) \exp[{i(\omega \tau - k \xi)}] d\xi d\tau
\end{equation}

Now ,
\begin{equation}
\widehat{(\frac{\partial f_{\epsilon}^{(1)}}{\partial \xi})}(\omega, k)=
(i k )\widehat{ f_{\epsilon}^{(1)}}(\omega, k),
\end{equation}
and
\begin{equation}
\widehat{(\frac{\partial f_{\epsilon}^{(1)}}{\partial \tau})}(\omega, k)=-(i \omega )\widehat{ f_{\epsilon}^{(1)}}
(\omega, k) -\frac{1}{2 \pi}\int_{\xi =-\infty}^{\infty} \exp[ -i k \xi]
f_{\epsilon}^{(1)}|_{\tau = 0}  d \xi ,
\end{equation}
and

\begin{equation}
 \widehat{\frac{\partial^3 \phi^{(1)}}{\partial \xi^3}}(\omega, k)=
(-i k^3 )\widehat{{ \phi^{(1)}}}(\omega, k)
\end{equation}
 Now applying these Fourier transforms on (\ref{uneque1}), letting $\epsilon \rightarrow 0$
and using 
\begin{equation}
  \lim_{\epsilon \to 0} \frac{1}{(k v - \omega \alpha_1 \epsilon^2)} = P(\frac{1}{k v}) + i \pi \delta(k v)
\end{equation}
we get,
\begin{equation}
 \widehat{ f^{(1)}}(\omega, k) = \widehat{ \phi^{(1)}} (\omega, k)f^{(0)}
 - Q k^2(3 - v^2)\widehat{ \phi^{(1)}}(\omega, k) f^{(0)} ,
\end{equation}
where $P$ is the principal part of the integral.
Taking inverse Fourier transform we get the form of $f^{(1)}$ as,
\begin{equation}
  f^{(1)} =  \phi^{(1)} f^{(0)} +Q (3 - v^2) \frac{\partial^2 \phi^{(1)}}{\partial \xi^2} f^{(0)} 
\label{f1}
\end{equation}
The first term of (\ref{f1}) is same with the classical case whereas the second term is the quantum correction
term.
Thus the procedure yields that $\lambda (\xi,\tau)$ appearing in (\ref{nonuneque}) is zero.

\subsection{$\epsilon^2$ order calculation:}
From equations (\ref{continuity})- (\ref{Poisson}), we can obtain in a straightforward way,

\begin{equation}
 2 \frac{\partial n^{(1)}}{\partial {\tau}} + 3 \alpha_2 n^{(1)}\frac{\partial n^{(1)}}{\partial {\xi}}+2 \alpha_3
\frac{\partial^3 n^{(1)}}{\partial \xi^3} = \alpha_2 \frac{\partial}{\partial \xi}(n_e^{(2)}-
 \phi^{(2)})
\label{AnotherE}
\end{equation}

From equation (\ref{Wigner}) we get,
\begin{equation}
 (\alpha_1 \epsilon^2)\frac{\partial f_{\epsilon}^{(2)}}{\partial {\tau}}
+ v \frac{\partial f_{\epsilon}^{(2)}}{\partial {\xi}} - v f^{(0)} \frac{\partial \phi^{(2)}}{\partial \xi}
- Q \frac{\partial^3 \phi^{(2)}}{\partial \xi ^3}\frac{\partial^3 f^{(0)}}{\partial v^3}
= C(\xi, \tau, v) ,\label{mainE1}
\end{equation}
where the $\tau$ derivative term is taken from $\epsilon^4 $ order and terms which are product of
quantum term and second order perturbation term are neglected as small compared to other terms.
Here $C(\xi, \tau, v)$ is defined as 
\begin{equation}
 C(\xi,\tau,v) = [C_a (\xi,\tau) + C_b (\xi,\tau) v + C_c(\xi, \tau)v^2 + C_d(\xi,\tau)v^3] f^{(0)},
\end{equation}
where
\begin{equation}
 C_a(\xi,\tau) = (\frac{\alpha_1}{\alpha_2})[\frac{\partial \phi^{(1)}}{\partial \xi} + 
3 Q \frac{\partial^3 \phi^{(1)}}{\partial \xi^3}]
\end{equation}

\begin{equation}
 C_b(\xi,\tau) = [\phi^{(1)} \frac{\partial \phi^{(1)}}{\partial \xi} + 
5 Q \frac{\partial^2 \phi^{(1)}}{\partial \xi^2}\frac{\partial \phi_1}{\partial \xi}
+ 3Q \phi^{(1)} \frac{\partial^3 \phi^{(1)}}{\partial \xi^3}]
\end{equation}

\begin{equation}
 C_c(\xi,\tau) = -Q (\frac{\alpha_1}{\alpha_2})[\frac{\partial^3 \phi^{(1)}}{\partial \xi^3} ]
\end{equation}

\begin{equation}
 C_d(\xi,\tau) = [
- Q \frac{\partial^2 \phi^{(1)}}{\partial \xi^2}\frac{\partial \phi_1}{\partial \xi}
-Q \phi^{(1)} \frac{\partial^3 \phi^{(1)}}{\partial \xi^3}]
\end{equation}

Introducing Fourier transform in (\ref{mainE1}) and letting $\epsilon$ tends to zero we get,
\begin{eqnarray}
 \widehat{f^{(2)}}(\omega,k) - f^{(0)} \widehat{\phi^{(2)}}(\omega,k)=
- i \widehat{C_a}[P(\frac{1}{k v})+ i \pi \delta(k v)] f^{(0)} - i v \widehat{C_b} P(\frac{1}{k v}) f^{(0)}\nonumber\\-
i v^2 \widehat{C_c} P(\frac{1}{k v}) f^{(0)}-i v^3 \widehat{C_d} P(\frac{1}{k v}) f^{(0)}
\end{eqnarray}

Multiplying by $(i k)$ and integrating over v yields
\begin{equation}
 i k \widehat{n^{(2)}} - i k \widehat{\phi^{(2)}} = i \sqrt{\frac{\pi}{2}}\widehat{C_a} sgn(k)
+ \widehat{C_b} + \widehat{C_d},
\label{n2phi2f}
\end{equation}
where we have used $k \delta(k v) = sgn(k) \delta(v)$.

Now taking inverse Fourier transform of equation (\ref{n2phi2f}) we obtain,
\begin{eqnarray}
 \frac{\partial}{\partial \xi}(n^{(2)} - \phi^{(2)}) = C_b + C_d -
\frac{1}{\sqrt{2 \pi}}[P \int_{-\infty}^{\infty}(\frac{\alpha_1}{\alpha_2})\frac{\partial n^{(1)}}{\partial \xi'}
\frac{d \xi'}{\xi - \xi'}+P \int_{-\infty}^{\infty}(\frac{ 3 Q \alpha_1}{\alpha_2})\frac{\partial^3 n^{(1)}}
{\partial \xi'^3}
\frac{d \xi'}{\xi - \xi'}],
\label{n2phi2}
\end{eqnarray}
Now using (\ref{AnotherE}) and (\ref{n2phi2}) we get finally,
\begin{eqnarray}
 \frac{\partial n^{(1)}}{\partial \tau} +
\alpha_{2} n^{(1)}\frac{\partial n^{(1)}}{\partial \xi}
+\alpha_{3}\frac{\partial^3 n^{(1)}}{\partial \xi^3} -                                   
Q \alpha_{2}\frac{\partial}{\partial \xi}
[\frac{\partial n^{(1)}}{\partial \xi}]^2 -
 Q \alpha_{2} n^{(1)} 
\frac{\partial^3 n^{(1)}}{\partial \xi^3}+
\nonumber\\
\frac{\alpha_{1}}{\sqrt{8\pi}}[P \int_{-\infty}^{\infty}\frac{\partial n^{(1)}}{\partial \xi'}\frac{d \xi'}{(\xi - \xi')}]
+\frac{3\alpha_{1} Q}{\sqrt{8\pi}}[P \int_{-\infty}^{\infty}\frac{\partial^3 n^{(1)}}{\partial \xi'^3}
\frac{d \xi'}{(\xi - \xi')}],
= 0 \nonumber\\
\label{FinalE}
\end{eqnarray}
 
which is the main equation of interest 
of this work. This equation implies the evolution equation of motion of nonlinear ion acoustic wave taking into account
the Landau damping effect with quantum corrections arising from semiclassical kinetic approach i.e, the
 Wigner equation approach.
The fourth and fifth terms of (\ref{FinalE})
 are nonlinear quantum corrections
 and the last term of the LHS is the quantum correction on the Landau damping. We can see
  that the equation converges exactly to the equation derived by 
Ott and Sudan {\cite{OttSudan1}} in the limit $\hbar$ $\rightarrow$ 0. The equation is like a higher order
KdV equation which have higher order nonlinear quantum correction terms and Landau damping term
with its quantum correction. Due to
the nature of the equation we can show that it conserves total number of particles. The presence of
Landau damping terms also assure that the amplitude of soliton must decay with time.
These relevant facts are derived in the next section.
%-------------------------------------------------------------------------------------------------------------------------
\section{Some relevant properties}

\subsection{Conservation of ion number }
The equation (\ref{FinalE}) is the higher order KdV equation with Landau damping terms.
Integrating (\ref{FinalE})w.r.to $\xi$ and assuming $n^{(1)}, {\partial n^{(1)}}/{\partial \xi},
{\partial^2 n^{(1)}}/{\partial \xi^2} = 0$ at $\xi = \pm \infty$ and renaming
$n^{(1)} = U $, we can show that  
\begin{equation}
\frac{\partial}{\partial \tau}  \int_{-\infty}^{\infty} U d\xi = 0                                
\end{equation}
Here we have used the fact that 
\begin{equation}
 P \int_{-\infty}^{\infty} \frac{d \xi}{\xi-\xi'} = 0
\end{equation}
Hence ion number  is conserved.
 \subsection{Decay of solitary wave}

Ott and Sudan in their paper {\cite{OttSudan1,OttSudan2}} considered $\alpha_1$ to be a small 
perturbation parameter and used the fact that due to Landau
damping the amplitude of KdV solitary wave will decrease with time. Then
using Bogoliubov- Mitropolsky {\cite{Bogoliubov}} approximation method they
found the decay rate of amplitude, which depends on the small parameter $\alpha_1$ . 
 In (\ref{FinalE}),we see that there are higher order KdV terms with Landau damping term and its quantum correction.
But since exact Sech- solitary wave solution of a general higher order KdV equation of above form is possible
only when (coefficient of the term $\frac{\partial U}{\partial \xi}
[\frac{\partial U}{\partial \xi}]^2$) = -2
 (coefficient of the term $ U 
\frac{\partial^3 U}{\partial \xi^3}$), which is not present in (\ref{FinalE}), hence the exact solitary wave
solution of the higher order KdV equation and its decay due to Landau damping terms cannot be worked out here.
Also it can be seen that (\ref{FinalE}) contains $2$ small parameters $\alpha_1$ and $Q$
 where $\alpha_2, \alpha_3$ are assumed to be $\approx 1$. Hence in the subsequent part of the work, the quantum correction
terms and the Landau damping term are treated as perturbation term to the KdV equation.
But since perturbation with multiple small parameters will include multiple time scales in the calculation,
hence it will be too complicated to be computed analytically.
%Since $\alpha_1 = \sqrt{m_e/m_i}$ is itself a small quantity
In order to simplify the case and find out the nature of
decay of the KdV solitary wave amplitude we will assume that $\alpha_1 \approx \alpha_2 Q$.
For example, in the case of hydrogen plasma $\alpha_1$ is approximately 0.025 and in {\cite{QElecHoles1,
QElecHoles2}}, 
the factor $Q$ is taken to be equal to be order of 
$ 0.01$. Assuming this relation between small parameters we can consider that
  the quantum correction to the Landau damping term
which appears as the last term of (\ref{FinalE}) is $\alpha_1\approx$ $Q^2$,   and
hence it can be neglected as small compared to the other terms.

Now we have to apply the 
well known method of Bogoliubov and Mitropolsky \cite{Bogoliubov,OttSudan1,OttSudan2} with $\alpha_2 Q
= C $ as small perturbation parameter.  Hence $\alpha_1$ can be taken as 
$\alpha_1 = \beta C$ where $\beta$ is any number $\approx$ unity. In order for the perturbation 
analysis to be consistent with the condition of validity of (\ref{FinalE}) it is also required that 
$1 \gg C \gg \epsilon$.
Assuming a new phase co-ordinate to have the form
\begin{equation}
 \phi(\xi, \tau) =\sqrt{\frac{N(\tau)\alpha_2}{12 \alpha_3}}(\xi - \frac{\alpha_2}{3}\int_{0}^{\tau}N(\tau) d\tau),
\label{phase}
\end{equation}
where $N(\tau)$ is assumed to vary slowly with time. 

We introduce two time scales
following \cite{Bogoliubov} as 
\begin{equation}
 t_0 = \tau,  t_1 = C \tau ,
\label{timescale}
\end{equation}
and $N = N(C, \tau)$
 and shall seek a solution 
of the form 
\begin{equation}
 U (\phi, C, \tau) = U_0(\phi, t_0, t_1) + O(C),
\label{amplexp}
\end{equation}
where (\ref{amplexp}) is to be valid for long times,i.e., times as large as $\tau \sim O(1/C)$.
In order to find such a solution, valid for long times, we first expand $u(\phi, \tau, C)$ to $O(C)$:
\begin{equation}
 U(\phi, \tau, C) = U_0(\phi, t_0, t_1) + C U_1(\phi, t_0) + O(C^2)
\label{amplexp2}
\end{equation}
Using (\ref{phase}), (\ref{timescale}), (\ref{amplexp2}) in (\ref{FinalE}) we get an equation containing
different powers of $C$ and equating coefficients of each power of $C$ we get different order equations
which need to be solved.

Since we are interested in the damping of solitary waves, we have the following initial and boundary conditions:
\begin{eqnarray}
 U(\phi, 0, C) = N_{0} sech^2(\phi), \nonumber\\
U (\pm{\infty}, \tau, C) = 0 
\label{IBc1}
\end{eqnarray}
Solving the order unity equation which is
\begin{equation}
 \rho \frac{\partial U_0}{\partial t_0} + \frac{\partial^3 U_0}{\partial \phi^3}
-4 \frac{\partial U_0}{\partial \phi} + \frac{12}{N} U_0 \frac{\partial U_0}{\partial \phi} = 0,
\label{unity}
\end{equation}
we get 
\begin{equation}
 U_{0} (\phi, t_0, t_1) = N(t_1)sech^2(\phi),
\end{equation}
where $\rho= {24 \sqrt{3\alpha_3}}/{(N \alpha_2)\sqrt{N \alpha_2}}$ and $N(t_1)$ is an arbitrary function of $t_1$
except for the initial condition $N(0) = N_0$. Hence $U_0$ doesn't depend on $t_0$.

The order $C$ equation is 
\begin{equation}
 \frac{\partial U_1}{\partial t_0} + L[U_1] = M[U_0],
\end{equation}
where
\begin{eqnarray}
 M[U_0] = -\frac{\partial U_0}{\partial t_1} - \frac{\phi}{2N}\frac{\partial U_0}{\partial \phi} \frac{dN}{dt_1}
+ \frac{1}{(\rho \alpha_3)}[ \frac{\partial^3 U_0}{\partial \phi^3} U_0
+2 \frac{\partial U_0}{\partial \phi} \frac{\partial^2 U_0}{\partial \phi^2}]
-\frac{\beta}{\sqrt{8 \pi}}[P \int_{-\infty}^{\infty}\sqrt{\frac{N(\tau)\alpha_2}{12 \alpha_3}}
\frac{\partial U_0}{\partial \phi'}
\frac{d \xi'}{\xi - \xi'}],
\label{Mu0}
\end{eqnarray}
\begin{eqnarray}
L[U_1] = \frac{1}{\rho}\frac{\partial^3 U_1}{\partial \phi^3} - \frac{4}{\rho}\frac{\partial U_1}{\partial \phi}
+ \frac{12}{(N \rho)}\frac{\partial (U_0 U_1)}{\partial \phi}
\end{eqnarray}
Again the boundary and initial conditions are
\begin{eqnarray}
 U_1 (\pm \infty, t_0) = 0,
U_1(\phi, 0) = 0
\label{IBc2}
\end{eqnarray}
In order that (\ref{amplexp2}) to be valid for times as large as $\tau \sim O(1/C)$
it is required that $U_1(\phi, t_0)$ does not behave secularly with $t_0$. To eliminate secular behavior
of $U_1$ it is necessary that $M[U_0]$ be orthogonal to all solutions, $g(\phi)$,
of $L^+[g] = 0$
which satisfy (\ref{IBc2})[i.e, $g(\pm \infty) = 0$], where $L^+$ is the operator adjoint to $L$ given by,
\begin{equation}
 L^+ = -\frac{1}{\rho}\frac{\partial^3}{\partial \phi^3} + \frac{4}{\rho}\frac{\partial}{\partial \phi}
-\frac{12}{\rho} sech^2(\phi)\frac{\partial}{\partial \phi}.
\end{equation}
The only solution of $L^+[g] = 0$, $g(\pm \infty) = 0$, is $g(\phi) = sech^2(\phi)$.

Thus,
\begin{equation}
 \int_{-\infty}^{\infty} sech^2(\phi) M[U_0] d\phi = 0
\label{OrthogC} 
\end{equation}

In order to evaluate this integral we have to consider term by term of (\ref{Mu0}). The first 2 terms of 
$M[U_0]$ together give $-{d N}/{d t_1}$ after integration.
The third and fourth terms which come from the nonlinear quantum correction terms give zero after
integration due to the odd nature of the integrand. Finally the last term, i.e the Landau damping term gives
 $(2.92)\frac{-\beta ({\alpha_2}}/{\sqrt{96 \pi \alpha_3}}) N\sqrt{N} $,
where we have used that 
\begin{equation}
 P \int_{-\infty}^{\infty} \int_{-\infty}^{\infty} sech^2(\phi) \frac{\partial(sech^2(\phi'))}{\partial \phi'}
 d\phi
\frac{d \phi'}{(\phi - \phi')} = (24/\pi^2) \zeta(3) = 2.92
\end{equation}
Thus we get a first order differential equation in $N$, solving which we get 
\begin{equation}
 N = \frac{N(0)}{[1 + (\frac{1}{2} \beta_1 \alpha_2 Q  N(0)^{(\frac{1}{2})}) \tau]^2},
\label{decayrate}
\end{equation}
where $\beta_1 = (2.92) \beta \sqrt{{\alpha_2}/{96 \pi \alpha_3}}$.

 From eqn (\ref{decayrate})
we see that the decay law of amplitude depends on the quantum factor $Q$. A full numerical computation of
(\ref{FinalE}) could reveal the total dynamical nature of the solution.

\begin{figure}[!h]
\centering

{
 \includegraphics[width=7 cm, angle=0]{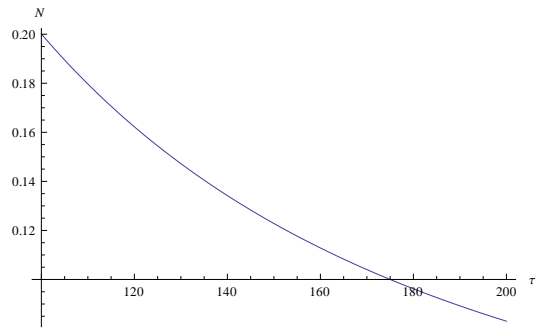}
}

\caption{Decay of soliton amplitude with time when when $Q =0.01 , N(0) = 1, \alpha_3=1,
\alpha_2=6, $
and $\beta=1$ 
}
\label{fig:1}
\end{figure}
\section{Concluding remarks:}

	In this work  we have extended the methodology of the work of
 Ott and Sudan to include the semiclassical quantum effects to obtain a new evolution equation in the context of  a nonlinear 
ion acoustic wave. This equation is of the form of a higher order KdV equation having higher order nonlinear terms as quantum corrections, together with a classical Landau damping term as well
as quantum contribution coming from resonant particle effects.

Using the fluid equations for ions and the classical  kinetic Vlasov equation for electrons, Ott and Sudan obtained a KdV equation with a Landau damping term as
the evolution equation for the nonlinear ion acoustic wave.
In order to introduce the quantum corrections, the classical Vlasov equation is  replaced
by an appropriate quantum analog i.e, the Wigner equation. 
In a similiar approach using the Wigner equation in place of the Vlasov equation gives rise to our higher order KdV equation with Landau damping terms.
 The equation exactly converges to the equation done in
 {\cite{OttSudan1}} when $\hbar$ tends to zero i.e, in the classical limit.  The mathematical nature of the equation shows that it conserves the total number of ions. The importance of the higher order KdV equation derived here,  lies  in the fact that its solution would give the quantum modification of the KdV solitary wave. But unfortunately, exact solitary wave solutions of this equation cannot be obtained.  Since there are  two small parameters in the equation,
 $\alpha_1$ and $Q$,  we treat the quantum corrections as well as the Landau damping terms as perturbation to the KdV equation. 
In order  to carry out the Bogoliubov and Mitropolsky approximation technique,  multiple time scales stretched by these small parameters have to be introduced.   Such a technique is too complicated to comprehend analytically.  Hence in order to get a
useful analytical result , we have assumed $\alpha_1 \approx Q$.
Hence, the quantum correction to Landau damping term turns out to be of the order of $Q^2$ and therefore neglected.  

In the perturbative approach, the contribution to the decay rate coming from the nonlinear quantum correction terms turns out to be zero because of the odd nature of the integrand. 
The final contribution to the decay of solitary wave amplitude comes from the classical Landau terms, whose coefficient, due to the perturbation scheme, turns out  to be of the order of $Q$. 
The amplitude is shown to decay inversely 
with the square of time 
depending on the quantum factor $Q$.  In our final equation
 of decay rate no terms come  from the quantum correction, i.e
  quantum nonlinear part goes to zero when the integration over $\phi$ is performed  and   
 the quantum Landau damping terms being of order $Q^2$
 are neglected. This is due to our chosen  scheme,
 and application of perturbation scheme with multiple time scales
 could give rise to solutions with more appropriate dependance
 on quantum effects. But the 
importance of the equation cannot be turned down and 
could be the initiator of  numerical computation that would reveal the entire dynamical nature of the solution with the inclusion of quantum mechanical effect.


\newpage

\begin{thebibliography}{99}

\bibitem{Astro} Y. D. Jung, Phys. Plasma {\bf{8}}, 3842 (2001)
\bibitem{Laser} D. Kremp, Th. Bornath, M. Bonitz, and M. Schlanges, Phys. Rev. E {\bf{60}}, 4725 (1999)
\bibitem{Electronics1} N. C.  Kluksdahl, A. M. Kriman, D. K. Ferry, and C. Ringhofer, Phys. Rev. B {\bf{39}}, 7720 (1989)
\bibitem{Electronics2} A. A. G. Driskill- Smith, D.G.Hasko and H.Ahmed, Appl. Phys. Lett {\bf{75}}, 2845 (1999)
\bibitem{Haas1} F. Haas, Quantum Plasmas- An hydrodynamic approach, Springer, New York (2011)
\bibitem{Manfredi} G. Manfredi and F. Haas, Phys. Rev. B 64, 075316 (2001).
\bibitem{Bonitz} M. Bonitz, AIP Conf. Proc.{\bf{1421}}, 135 (2012)
\bibitem{Haas2} F. Haas, B. Eliasson, P. K. Shukla, and G. Manfredi,
Phys. Rev.  E 78, 056407 (2008).
\bibitem{Gardner}C.L Gardner, SIAM. J.Appl. Math {\bf{54}}, 2(409).
\bibitem{Lange} H. Lange, B. Toomire and P.F. Zweifel, Trans. theor. Stat .Phys. {\bf{25(6)}}, 713 (1996).
\bibitem{OttSudan1} E. Ott  and R.N. Sudan,  Phys. Fluids {\bf{12}}, 11(1969).
\bibitem{AnupB}  A. Bandyopadhyay and K.P. Das, Phys. Plasma, {\bf{9}}, 2 (2002).
\bibitem{QElecHoles1} A. Luque, H.  Schamel, and R.  Fedele Phys. Lett. A {\bf 324}, 185-192 (2004)
\bibitem{QElecHoles2} D. Jovanovic, R. Fedele, Phys. Lett. A {\bf 364}, 304-312 (2007)
\bibitem{Bogoliubov}  Bogoliubov N. N,  Mitropolsky Y. A, Asymptotic Methods in the Theory of Nonlinear Oscillations
(Gordon and Breach Science Publishers, Inc, New York, 1961)

\bibitem{OttSudan2} Ott E, Sudan R. N  1970 {\it Phys. Fluids} {\bf 13} 6
\bibitem{Salimullah} M. Salimullah, M.  Jamil, I.  Zeba, Ch.  Uzma, and H. A.  Shah, Phys. Plasma. {\bf{16}}, 034503(2009)
\bibitem{Haijun} H. Ren, Z.  Wu, J.  Cao, and P. K.  Chu, J. Phys. A. {\bf{41}}, 115501(2008).


%M. Lakshmanan  and S. Rajasekar, {\it Nonlinear dynamics: Integrability, Chaos and Patterns},
%(Springer, Berlin). 2003

 \end{thebibliography}
\end{document}